\documentclass[letterpaper, 10 pt, conference]{ieeeconf}  
\usepackage{graphics} 
\usepackage{epsfig} 
\usepackage{times} 
\usepackage[utf8]{inputenc}
\usepackage[english]{babel}
\usepackage{amsmath}
\usepackage{amssymb}
\usepackage{hyperref} 
\usepackage{rotating}
\usepackage{graphicx}
\usepackage{caption}
\usepackage{subcaption}
\usepackage{tikz}
\usepackage{dirtytalk}
\usetikzlibrary{shapes.geometric, arrows}

\newcommand{\ols}[1]{\mskip.5\thinmuskip\overline{\mskip-.5\thinmuskip {#1} \mskip-.5\thinmuskip}\mskip.5\thinmuskip} 
\newcommand{\olsi}[1]{\,\overline{\!{#1}}} 
\makeatletter
\newcommand\closure[1]{
  \tctestifnum{\count@stringtoks{#1}>1} 
  {\ols{#1}} 
  {\olsi{#1}} 
}
\long\def\count@stringtoks#1{\tc@earg\count@toks{\string#1}}
\long\def\count@toks#1{\the\numexpr-1\count@@toks#1.\tc@endcnt}
\long\def\count@@toks#1#2\tc@endcnt{+1\tc@ifempty{#2}{\relax}{\count@@toks#2\tc@endcnt}}
\def\tc@ifempty#1{\tc@testxifx{\expandafter\relax\detokenize{#1}\relax}}
\long\def\tc@earg#1#2{\expandafter#1\expandafter{#2}}
\long\def\tctestifnum#1{\tctestifcon{\ifnum#1\relax}}
\long\def\tctestifcon#1{#1\expandafter\tc@exfirst\else\expandafter\tc@exsecond\fi}
\long\def\tc@testxifx{\tc@earg\tctestifx}
\long\def\tctestifx#1{\tctestifcon{\ifx#1}}
\long\def\tc@exfirst#1#2{#1}
\long\def\tc@exsecond#1#2{#2}
\makeatother

\title{\LARGE \bf
Online Adaptive Identification of Switched Affine Systems Using a Two-Tier Filter Architecture with Memory}
\author{Pritesh Patel$^{1}$, Sayan Basu Roy$^{2}$ and Shubhendu Bhasin$^{3}$
\thanks{$^{1}$Pritesh Patel is a PhD student in Control and Automation, Electrical Engineering, Indian Institute of Technology Delhi, India
        {\tt\small priteshpatel.iitd@gmail.com}}%
\thanks{$^{2}$Sayan Basu Roy is Assistant Professor at  Indraprastha Institute of Information Technology Delhi, India
        {\tt\small sayan@iiitd.ac.in}}%
\thanks{$^{3}$Shubhendu Bhasin is Associate Professor at  Indian Institute of Technology Delhi, India
        {\tt\small sbhasin@ee.iitd.ac.in}}%
}
\begin{document}
\maketitle
\thispagestyle{empty}
\pagestyle{empty}

\begin{abstract}
This work proposes an online adaptive identification method for multi-input multi-output (MIMO)  switched affine systems with guaranteed parameter convergence. A family of online parameter estimators is used that is equipped with a dual-layer low pass filter architecture to facilitate parameter learning and identification of each subsystem. The filters capture information about the unknown parameters in the form of a prediction error which is used in the parameter estimation algorithm. A salient feature of the proposed method that distinguishes it from most previous results is the use of a memory bank that stores filter values and promotes parameter learning during both active and inactive phases of a subsystem. Specifically, the learnt experience from the previous active phase of a subsystem is retained in the memory and leveraged for parameter learning in its subsequent active and inactive phases. Further, a new notion of intermittent initial excitation (IIE) is introduced that extends the previously established initial excitation (IE) condition to the switched system framework. IIE is shown to be sufficient to ensure exponential convergence of the switched system parameters.

\end{abstract}

\section{INTRODUCTION}
 Hybrid systems are generally characterized by co-operation between continuous and discrete dynamics in the sense that continuous dynamics takes values from discrete states as input and vice-versa, e.g. automotive systems, communication networks, energy systems, etc. Switched systems are a sub-class of hybrid systems with discrete switching events. Adaptive identification of switched affine systems involves online estimation of plant parameters of the each subsystem of the switched system \cite{liberzon2003switching}, \cite{garulli2012survey} (see also references therein). 
 
 In classical adaptive systems, which do not involve switching, a single parameter estimator is sufficient for estimating the unknown parameter vector. However, in the case of switched systems, where multiple subsystems switch between each other, a single estimator may not be effective due to the transient period associated with learning a new parameter, after every switching event. For safety-critical applications, e.g., for a pick and place robot, where the system parameters are different for different payloads, reduced transients after every switching are desirable to improve overall system performance. Online identification algorithms for piecewise affine (PWA) systems, with and without the knowledge of the switching signal are presented in \cite{kersting2014online} and \cite{kersting2014adaptive}, respectively. Both methods are extension of
the well-studied series-parallel parameter identifiers in adaptive
control to PWA systems. Parameter estimation of state-dependent switched system and model reference adaptive control (MRAC) for switched systems is developed using the concurrent learning approach in \cite{kersting2014concurrent} and \cite{de2013concurrent}, respectively. Both of these works assume that the switching signal is known, and it is shown that linear independence of the recorded data is sufficient for parameter convergence. 
In \cite{vidal2008recursive} subsystem models of deterministic discrete-time Switched Auto-Regressive eXogenous (SARX) are identified by assuming the number of models, the model order and switching signal being unknown. Identification of piecewise linear (PWL) dynamical systems and MRAC for PWA continuous system using minimal control synthesis algorithm is presented in \cite{di2009hybrid} and \cite{di2008novel}, respectively. Extensive literature for switched, PWA and PWL system identification is found in \cite{garulli2012survey} and references therein.
Three major limitations observed in the above literature are: (i) the parameter learning stops during the inactive period of the subsystem \cite{kersting2014online}, \cite{kersting2014adaptive}, \cite{yuan2016adaptive}, (ii) the requirement of intelligent-recording of large amount of data while a subsystem is in the active period \cite{kersting2014concurrent}, \cite{de2013concurrent}, \cite{chowdhary2010concurrent}; as the number of subsystems and the number of data points required to estimate the subsystem increases, computational complexity also increases and (iii) the requirement of persistence of excitation (PE) condition on the regressor for parameter convergence \cite{kersting2014online}, \cite{kersting2014adaptive}, \cite{vidal2008recursive},
\cite{yuan2016adaptive},
\cite{verdult2004subspace}, \cite{van2016online}.

The information gained about the unknown parameters during the active phase of a subsystem is typically lost due to switching to another subsystem. Hence, a major challenge in parameter estimation of switched systems is to devise a way to retain this information for later use.
The proposed work presents an online parameter estimation architecture for switched systems, where information about the unknown subsystem parameter, captured during the active phase of the subsystem, is stored in the memory in the form of filter outputs. The stored information is used for parameter learning in the subsequent inactive as well as active periods. The values of the filter variables, at the switch-out instants of a subsystem,  are required to be stored to continue parameter learning in the inactive period, unlike \cite{kersting2014concurrent}, where intelligent-recording of several independent data points is required in the active period to sustain parameter learning during the inactive period. Compared to the design in \cite{roy2017uges}, a different filter structure is proposed, where filter values at the switch-out instants are stored in a memory stack and are used to reset the values of filter variables at every switching instant. Another contribution of this result is the introduction of the notion of intermittent initial excitation (IIE), an extension of  previous works on initial excitation (IE) \cite{roy2017uges}, \cite{roy2016parameter}, that is shown to be sufficient to ensure parameter convergence for switched affine systems. Here, intermittent implies that a subsystem experiences both active and inactive phases during the course of operation, however, new information about the unknown parameters can be gleaned only during the active period. The IIE condition stipulates sufficient excitation during such intermittent periods when a subsystem is ON, for acquiring information about the unknown parameters. To capture the information rich data corresponding to a subsystem, dual layer low pass filters and their corresponding memory stacks are used, which store filter values for a subsystem at their switch-out instants. Parameter convergence is especially critical in the context of hybrid systems because any degradation in the system performance can lead to overall system instability due to the interconnected architecture \cite{yuan2016adaptive}. The proposed method guarantees uniform global exponential stability (UGES) of the overall parameter estimation error dynamics without the PE condition; the milder IIE
condition for each subsystem is imposed on the regressor to ensure parameter convergence.   \\
\textit{Notations}: $\mathbb{R}^n$ denotes the real $n$-vector; $\mathbb{R}^{n\times m}$ denotes the real $n\times m$ matrices; $\mathbb{N}$ denotes the set of natural numbers; $||\bullet||$ denotes the Euclidean norm of a vector; $tr\{\bullet\}$ is the trace of a matrix; $I_n$ denotes the identity matrix of order $n$; \say{$\otimes$} denotes the matrix Kronecker product; $vec(Z)\in\mathbb{R}^{ab}$ denotes the vectorization of a matrix $Z\in\mathbb{R}^{a\times b}$ obtained by stacking the columns of the matrix $Z$.


\section{PROBLEM FORMULATION AND PRELIMINARIES}\label{s2}
\subsection{System description}
 Consider the following uncertain switched linear time invariant (LTI) system
 \begin{equation}\label{eq1}
     \dot{x}(t)=A_{\sigma(t)}x(t)+B_{\sigma(t)}u(t), \hspace{0.3cm}x(t_0)=x_0 
 \end{equation}
where $x(t)\in \mathbb{R}^n$ is the state vector, $u(t)\in \mathbb{R}^m$ is the control input, $t\in[t_0,\infty),\hspace{0.2cm} t_0\geq 0$, $\sigma : [0, \infty) \rightarrow \mathbf{S}$ denotes a piecewise constant switching signal, where $\mathbf{S}=\{1, 2, 3,...,M\}$,  $A_{i}\in\mathbb{R}^{n\times n}$ \& $B_{i} \in \mathbb{R}^{n \times m}$ denotes system and input matrices respectively for the $i^{th}$ subsystem ($i\in \mathbf{S}$). The system starts from an initial time $t_0$, and let $t_k\hspace{0.1cm}(\forall k \in \mathbb{N})$ denote the time instants when the system switches from one subsystem to another based on the switching signal $\sigma(t)$, which is discontinuous at the switching instants and has a constant value between two consecutive switching instants. 
   At each time instant, $\sigma(t)$ specifies the index of the active subsystem from the family $\mathbf{S}$. Although the switching signal $\sigma(t)$ is not known a priori, its instantaneous value is assumed to be known at the current time instant. It is assumed that there are no discontinuous jumps in the state at the time of switching from one subsystem to another.
   
  The plant dynamics\footnote{A linear switched system (\ref{eq1}) is considered in this paper, however the development can be trivially extended to linearly parametrizable nonlinear systems that can be cast in the same form as (\ref{eq2})} in (\ref{eq1}) can be linearly parametrized as
   \begin{equation}\label{eq2}
       \dot{x}=Y(x,t)\theta_{\sigma(t)} \hspace{1cm} \sigma(t)\in \mathbf{S}   \end{equation}
where $Y(x,t) \in \mathbb{R}^{n \times n(n+m)}$ is a known regressor matrix, defined as
\begin{equation}
    Y\triangleq[X \hspace{0.2cm} U]
\end{equation}
where $X\in \mathbb{R}^{n\times n^2}$ and $U \in \mathbb{R}^{n \times nm}$ are given by
\begin{equation}
    X=I_n \otimes x^T, U=I_n \otimes u^T
\end{equation}
where \say{$\otimes$} denotes the matrix Kronecker product. The unknown switched parameter vector $\theta_{\sigma(t)}\in \mathbb{R}^{n(n+m)}$, containing all the elements of $A_{\sigma(t)}$ and $B_{\sigma(t)}$ is defined as
\begin{equation}\label{eq3}
    \theta_{\sigma(t)}\triangleq \begin{bmatrix}vec(A_{\sigma(t)}^T)\\
    vec(B_{\sigma(t)}^T)
    \end{bmatrix}
\end{equation}
where $vec(Z)\in \mathbb{R}^{ab}$ denotes the vectorization of a matrix $Z \in \mathbb{R}^{a\times b}$, obtained by stacking the columns of the matrix $Z$.
\subsection{Identification Objective}
The objective  is to design an parameter estimation law for each subsystem such that $\hat{\theta}_{i}(t) \rightarrow \theta_{i}$ as $t \rightarrow \infty$, $i\in \mathbf{S}$.\\
\subsection{Preliminary Definitions}

\textit{\textbf{Definition 1}} : A signal $\varphi(t) \in \mathbb{R}^{n \times m}$, where $m>n>0$ is persistently exciting (PE) w.r.t $\dot{x}=f(t,x)$ if $\forall\left(t_{0}, x_{0}\right) \in \mathbb{R}_{\geq 0} \times \mathbb{R}^{n}$, $\exists \alpha, T>0$ such that:
$$
\int_{t}^{t+T} \varphi^T(\tau) \varphi(\tau) d \tau \geq \alpha I_{m}, \quad \forall t \geq t_{0}
$$
where $T$ is the window-length of integration, and $\alpha$ is called the degree-of-excitation.\\
\textit{\textbf{Definition 2}}: A signal $\varphi(t, x) \in \mathbb{R}^{n \times m}$, where $m>n>0$, is called initially exciting (IE) w.r.t $\dot{x}=f(t,x)$ if $\forall\left(t_{0}, x_{0}\right) \in \mathbb{R}_{\geq 0} \times \mathbb{R}^{n}$, $\exists \alpha, T>0$, such that all corresponding solutions satisfy
$$
\int_{t_{0}}^{t_{0}+T} \varphi^T\left(s, x\left(s, t_{0}, x_{0}\right)\right) \varphi\left(s, x\left(s, t_{0}, x_{0}\right)\right) d s \geq \alpha I_{m}
$$
where $T$ is the window-length of integration, and $\alpha$ is called the degree-of-excitation.
\\
\textit{\textbf{Definition 3}} : A signal $\varphi(t,x)\in \mathbb{R}^{n\times m}$, where $m>n>0$, is called \textit{intermittent} IE (IIE) w.r.t. $\dot{x}=f(t,x)$ and indicator function $\mathfrak{I}(t) :[0,\infty)\rightarrow \{0,1\}$ if $\forall (t_0,x_0)\in \mathbb{R}_{\geq 0}\times \mathbb{R}^n $, $\hspace{0.1cm} \exists \hspace{0.1cm}\alpha, T>0$, such that, all corresponding solutions satisfy
\begin{align}\label{eq15p}
   & \int_{t_{0}}^{t_{0}+T}\mathfrak{I}(t)\varphi^T(\tau, x(\tau,t_0,x_0))\varphi(\tau,x(\tau,t_0,x_0))d\tau\geq \alpha I_m,\nonumber\\
\end{align}
where the indicator function $\mathfrak{I}(t)\in \{0,1\}$ is a logic signal with a value of either 0 or 1.\\
\textit{\textbf{Remark 1}}: The IE condition in \cite{roy2017uges}, \cite{roy2017combined} is not directly applicable in the switched system context where subsystems undergo intermittent active and inactive phases, as dictated by the switching signal. The challenge is to capture rich information, available only during the active periods of a subsystem, and leverage it for parameter learning. The IIE condition proposed in this work uses indicator functions to stitch together the active periods of a subsystem. We show that IIE is sufficient for parameter convergence of switched affine systems, and therefore, generalizes the notion of IE.\\

\section{SWITCHED ADAPTIVE ESTIMATOR DESIGN}
\subsection{First Layer Filters}
Consider the following filter equation $\forall k \in \mathbb{N} \hspace{0.1cm} \& \hspace{0.1cm} \sigma(t)\in \mathbf{S}$
\begin{subequations}
\begin{align}
&\dot{N}(t)=-{k_f}N(t)+Y(x,t),  & &N(t_0)=0 \label{eq4} \\
&N(t_k)=\mathbf{S_{N_{\sigma}}}_{(t_k)} \label{eq4p}\\
&\dot{g}(t)= -k_fg(t)+\dot{x}(t),  & &g(t_0)=0 \label{eq6}\\
&g(t_k)=\mathbf{S_{g_{\sigma}}}_{(t_k)}
\label{eq6p}\\
&\dot{h}(t)=-k_fh(t)+x(t), & & h(t_0)=0 \label{eq_6e}\\
&h(t_k)=\mathbf{S_{h_{\sigma}}}_{(t_k)}
\label{eq_6ep}&
\end{align}
\end{subequations}
where $N(t)\in \mathbb{R}^{n\times n(n+m)}$ is the filtered regressor matrix, $g(t)\in \mathbb{R}^n$ denotes the filtered state derivative, $h(t)\in \mathbb{R}^n$ is the filtered state and $k_f>0$ is a scalar gain introduced to stabilize the filters. Further, $\mathbf{S_{N}}_i \in \mathbb{R}^{n\times n(n+m)}$, $\mathbf{S_{g}}_i \in \mathbb{R}^{n}$, $\mathbf{S_{h}}_i \in \mathbb{R}^{n}$
 denote the $i^{th}$ element of the memory stacks $\mathbf{S_{N}}$, $ \mathbf{S_{g}}$ and $\mathbf{S_{h}}$ respectively, which store the filter values at the switch-out instants corresponding to the $i^{th}$ subsystem ($i \in \mathbf{S})$. The memory stacks are defined as $\mathbf{S_{N}}\triangleq[\mathbf{S_{N_1}},\mathbf{S_{N_2}},...\mathbf{S_{N_M}}]$, $\mathbf{S_{g}}\triangleq [\mathbf{S_{g_1}},\mathbf{S_{g_2}},...\mathbf{S_{g_M}}]$, $\mathbf{S_{h}}\triangleq [\mathbf{S_{h_1}},\mathbf{S_{h_2}},...\mathbf{S_{h_M}}]$ and are initialized to zero, i.e. $\mathbf{S_{N}}_i(t_0)=0,\hspace{0.1cm}\mathbf{S_{g}}_i(t_0)=0, \hspace{0.1cm}\mathbf{S_{h}}_i(t_0)=0, \hspace{0.1cm}i\in \mathbf{S}$. The memory stacks are populated using the following logic
\begin{subequations}
\begin{align}
    &\mathbf{S_{N_{\sigma}}}_{(t_k^-)}={N}(t^-_k), \hspace{0.5cm}  \forall k \in \mathbb{N}&\\
&\mathbf{S_{g_{\sigma}}}_{(t_k^-)}={g}(t^-_k), \hspace{0.7cm} \forall k \in \mathbb{N}&\\
&\mathbf{S_{h_{\sigma}}}_{(t_k^-)}={h}(t^-_k), \hspace{0.7cm} \forall k \in \mathbb{N}&
\end{align}
\end{subequations}
where $t_k\hspace{0.1cm}(\forall k \in \mathbb{N})$ denotes the switching instant \& $t^-_k$ denotes the time just before the switching instant $t_k$.\\
\textit{\textbf{Remark 2}}: The equations (\ref{eq4}), (\ref{eq6}) and (\ref{eq_6e}) represent the continuous filter dynamics whereas (\ref{eq4p}), (\ref{eq6p}) and (\ref{eq_6ep}) indicate the discrete reset of filter states at switching instants $t_k\hspace{0.1cm}(\forall k \in \mathbb{N})$. At each switching instant $t_k$, when the system switches from, say, subsystem $q$ to subsystem $r\hspace{0.1cm} (q,r\in\mathbf{S}),$ the filter states $N(t_k^-),g(t_k^-)$ and $h(t_k^-)$ corresponding to the unknown parameter of subsystem $q$ are recorded in the memory stack at locations $\mathbf{S_{N}}_q$, $\mathbf{S_{g}}_q$ and $\mathbf{S_{h}}_q$ respectively. These stored filter values are later recalled whenever subsystem $q$ is switched back ON.\\
Filter equations (\ref{eq4}) and (\ref{eq6}) can be solved explicitly as
\begin{subequations}
\begin{align}
    &N(t)=N(t_k)+\exp\{-k_ft\}\int_{t_{k}}^{t}\exp\{k_f\tau\}Y(x(\tau),\tau)d\tau \label{eq8}&\\
    &N(t_k)=\mathbf{S_{N_{\sigma}}}_{(t_k)},\hspace{0,2cm}t\in [t_k,t_{k+1}), \forall k \in \mathbb{N}&
\end{align}
\end{subequations}
\begin{subequations}
\begin{align}
    &g(t)=g(t_k)+\exp\{-k_ft\}\int_{t_{k}}^{t}\exp\{k_f\tau\}\dot{x}(\tau) d\tau\label{eq9}&\\
    &g(t_k)=\mathbf{S_{g_{\sigma}}}_{(t_k)},\hspace{0.2cm}t\in [t_k,t_{k+1}), \forall k \in \mathbb{N}&
\end{align}
\end{subequations}
Substituting (\ref{eq2}) in (\ref{eq9}) and using (\ref{eq8}), the following relation is deduced:
\begin{equation}\label{eq10}
    g(t)=N(t)\theta_{\sigma(t)}, \hspace{0.2cm}  \forall t\geq t_0
\end{equation}
While $N(t)$ can be computed online using (\ref{eq4}), $g(t)$ cannot be solved from (\ref{eq6}) since $\dot{x}(t)$ is unknown. However (\ref{eq9}) can be further modified using the by parts rule of integration:
\begin{equation}
    g(t)=g(t_k)+x(t)-\exp\{-k_f(t-t_k)\}x(t_k)-k_f(h(t)-h(t_k))
\end{equation}
\hspace{4cm}$t\in [t_k,t_{k+1}), \forall k \in \mathbb{N}$\\   
where $h(t)$ can be computed from (\ref{eq_6e}).
Hence, state-derivative information is obviated in the proposed framework, unlike \cite{kersting2014concurrent}.
\subsection{Second Layer Filters}
Consider the following second layer of filter equations $\forall k \in \mathbb{N} \hspace{0.1cm} \& \hspace{0.1cm} \sigma(t)\in \mathbf{S}$
\begin{subequations}
\begin{align}
&\dot{Q}(t)=-k_sQ(t)+N(t)^TN(t),  &&Q(t_0)=0 \label{eq11}\\
&Q(t_k)=\mathbf{S_{Q_{\sigma}}}_{(t_k)}\\
&\dot{G}(t)=-k_sG(t)+N(t)^Tg(t), && G(t_0)=0\\
&G(t_k)=\mathbf{S_{G_{\sigma}}}_{(t_k)} 
\end{align}
\end{subequations}
where $k_s>0$ is a scalar gain and $Q(t)\in \mathbb{R}^{n(n+m)\times n(n+m)}$ denotes the double-filtered regressor and $G(t)\in \mathbb{R}^{n(n+m)}$. $\mathbf{S_{Q}}_i \in \mathbb{R}^{n\times n(n+m)}$, $\mathbf{S_{G}}_i \in \mathbb{R}^{n}$
 denote the $i^{th}$ element of memory stack $\mathbf{S_{Q}}$ and $ \mathbf{S_{G}}$, which store the filter value at the switch-out instants corresponding to the $i^{th}$ subsystem ($i \in \mathbf{S})$. The resulting memory stacks are defined as $\mathbf{S_{Q}}=[\mathbf{S_{Q_1}},\mathbf{S_{Q_2}},...\mathbf{S_{Q_M}}]$, $\mathbf{S_{G}}=[\mathbf{S_{G_1}},\mathbf{S_{G_2}},...\mathbf{S_{G_M}}]$ and are initialized to zero, i.e. $\mathbf{S_{Q}}_i(t_0)=0,\hspace{0.1cm}\mathbf{S_{G}}_i(t_0)=0,\hspace{0.1cm}i\in \mathbf{S}$. The memory stacks are populated using the following logic
\begin{subequations}
\begin{align}
    \mathbf{S_{Q_{\sigma}}}_{(t_k^-)}&={Q}(t^-_k), \hspace{0.5cm} \forall k \in \mathbb{N}\\
    \mathbf{S_{G_{\sigma}}}_{(t_k^-)}&={G}(t^-_k), \hspace{0.5cm} \forall k \in \mathbb{N}
    \end{align}
\end{subequations}
\textit{\textbf{Remark 3}}: The filters $N(t),g(t),Q(t),G(t)$ carry information about the unknown parameters for the active subsystem. To avoid information loss when the subsystem becomes inactive, the filter values at the switch-out instant are stored in a memory stack, only to be recalled when the subsystem becomes active again.\\
The following relation can be deduced in a similar way to (\ref{eq10}) as
\begin{equation}
    G(t)=Q(t)\theta_{\sigma(t)}, \hspace{0.8cm} \forall t\geq t_0\end{equation}
From (\ref{eq11}), the square matrix $Q(t)$ can be expressed as
\begin{subequations}
\begin{align}\label{eq28}
    &Q(t)=\underbrace{Q(t_k)}_{\geq 0}+\underbrace{\exp\{-k_st\}}_{\geq 0}\int_{t_{k}}^{t}\underbrace{\exp\{k_sr\}}_{\geq1}\underbrace{N(r)^TN(r)}_{\geq 0}dr& \\
    &Q(t_k)=\mathbf{S_{Q_{\sigma}}}_{(t_k)},t\in[t_k,t_{k+1}),\hspace{0.2cm} \forall k \in \mathbb{N}&
\end{align}
\end{subequations}
Using (\ref{eq28}), the following property can be derived.\\
\textit{\textbf{Property 1}}. $Q(t)$ is a positive semi-definite function of time i.e. $Q(t)\geq 0, \hspace{0.2cm} \forall t\geq t_0.$\\
\textit{\textbf{Assumption 1}}: The regressor $N(t,x)$ is IIE w.r.t. subsystem $i\hspace{0.1cm}(i\in\mathbf{S})$ in (\ref{eq1}) and indicator function $\mathfrak{I}_i(t):[0,\infty)\rightarrow \{0,1\}$ (as per Definition 1) with degree of excitation $\gamma_i$, i.e.,  
\begin{equation}\label{eq15}
    \int_{t_{0}}^{t_{0}+T_i}\mathfrak{I}_i(t)N(\tau, x(\tau))^TN(\tau,x(\tau))d\tau\geq \gamma_i I_{n(n+m)},  \hspace{0.1cm} i\in \mathbf{S}
\end{equation}
where $\mathfrak{I}_i(t)$ is an indicator function for the $i^{th}$ subsystem defined as
\begin{equation}
    \mathfrak{I}_i(t)=\begin{cases}
    1, & \text{when} \hspace{0.2cm}\sigma(t)=i\\
    0, & \text{when} \hspace{0.2cm} \sigma(t)\neq i
    \end{cases}
\end{equation}
\textit{\textbf{Remark 4}}: The IIE condition demands that the regressor have sufficient energy/richness in the initial time-window of activation time and it can extend to multiple active windows until the IIE condition of each subsystem is met. IIE is a more generalized version of IE \cite{roy2017uges, roy2017combined} and significantly less restrictive than PE \cite{narendra2012stable}, where the signal is required to have sufficient energy for the entire time-span.\\
\textit{\textbf{Remark 5}}: Assumption 1 states the existence of $T_i$, the time taken to satisfy IIE for each subsystem, implying that each subsystem is active often enough and the data $N(t)$ during the active period is rich enough such that the  IIE condition in (\ref{eq15}) is satisfied eventually.
All subsystems share the same filter variable $N(t)$; hence the use of indicator function as a multiplying factor in (\ref{eq15}) is done to ensure that the integral evaluates to zero during the inactive phase of a subsystem.
\begin{figure}
\includegraphics[width=9cm, height=5cm]{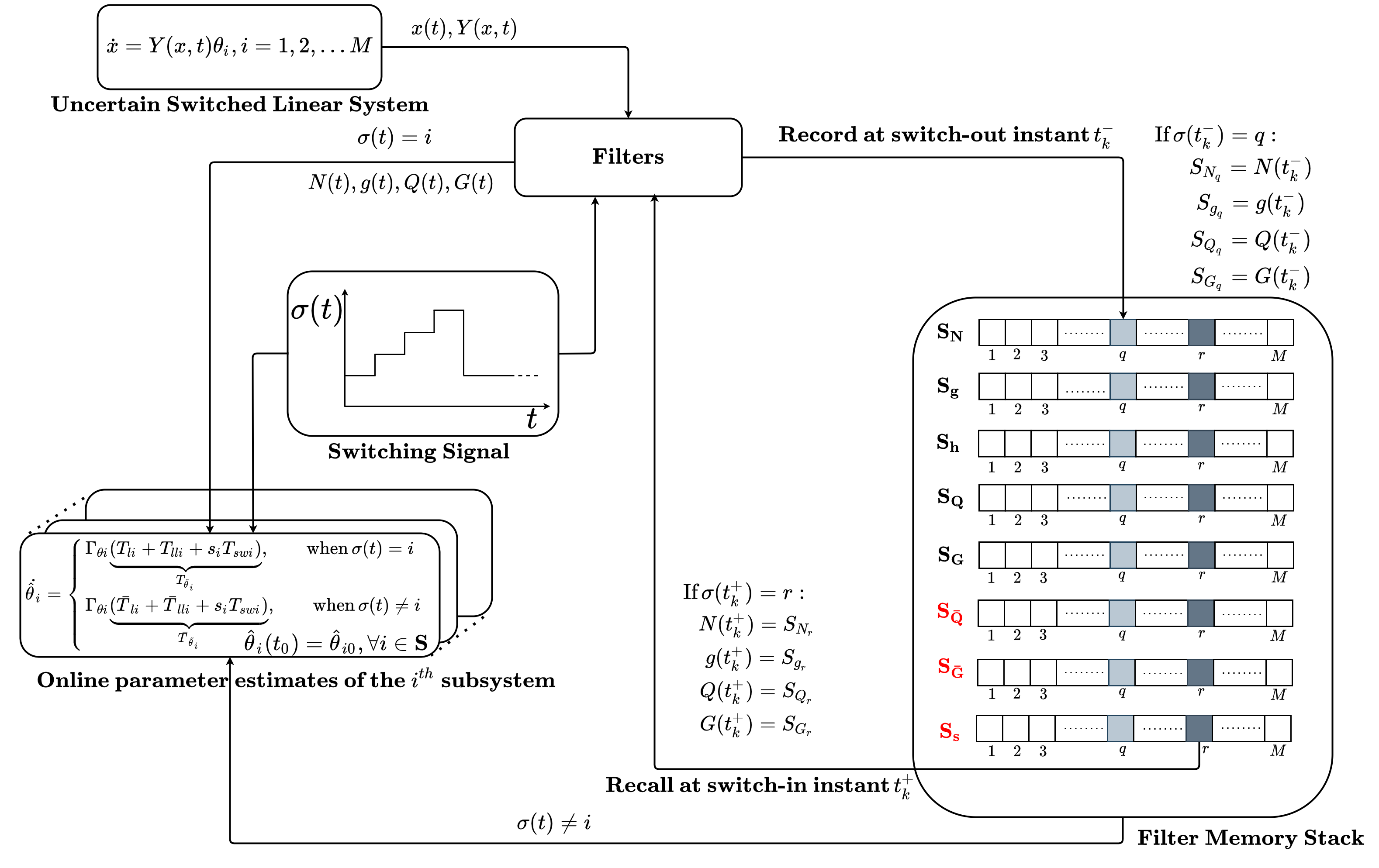}
\caption{Information flow block diagram for adaptive identification of uncertain switched LTI system}
\label{fig:my_label}
\end{figure}
\textit{\textbf{Lemma 1}}: A necessary and sufficient condition for the regressor $N(t,x)$ to be IIE for subsystem $i$ is that $Q(t_{0}+T_i)$ is a positive definite (PD) matrix, where $t_0+T_i \in \bar{t}_i$ and $\bar{t}_i=\{t\hspace{0.1cm} |\hspace{0.1cm} \sigma(t)=i\}$.\\
\textbf{Proof} : Consider the dynamics
\begin{equation}\label{eql1}
    \dot{N}_i(t)=\mathfrak{I}_i(t) N(t)^TN(t),\hspace{0.5cm} N_i(t_{0})=0, \hspace{0.5cm} \forall i \in \mathbf{S}  
\end{equation}
where $N_i(t)\in \mathbb{R}^{n(n+m)\times n(n+m)}$\\
Equation (\ref{eql1}) can be solved as
\begin{equation}\label{eql4}
       N_i(t)= \int_{t_0}^{t}\mathfrak{I}_i(\tau) N^T(\tau) N(\tau) d\tau, \hspace{0.2cm}   \forall t\geq t_0
\end{equation}
Let $Q_i(t)$ denote the second-layer filter state that captures the value of $Q(t)$(from \ref{eq11}) when subsystem $i$ is active
\begin{equation}\label{eql2}
    \dot{Q}_i(t)=-k_s\mathfrak{I}_i(t)Q_i(t)+\mathfrak{I}_i(t)N(t)^TN(t)
\end{equation}
$$\hspace{5cm}Q_i(t_0)=0, \hspace{0.2cm} \forall t\geq t_0$$
As $\mathfrak{I}_i(t)$ can take only constant values, equation (\ref{eql2}) can be solved as
\begin{equation}
    Q_i(t)=\exp\{-k_s\mathfrak{I}_it\}\int_{t_0}^{t}\exp\{k_s\mathfrak{I}_i\tau\}\mathfrak{I}_i N^TN d\tau, \hspace{0.2cm} \forall t\geq t_0
\end{equation}
$Q_i(t)$ can be upper and lower-bounded for the $i^{th}$ subsystem as
\begin{equation}
   Q_i(t) \leq \int_{t_0}^{t}\mathfrak{I}_iN^TN d\tau, \hspace{0.2cm} \forall t
\end{equation}
\begin{equation}
     Q_i(t) \geq \exp\{-k_s  (t-t_0)\mathfrak{I}_i\}\int_{t_0}^{t}\mathfrak{I}_iN^TN d\tau, \hspace{0.2cm} \forall t
\end{equation}
\begin{equation}\label{l34}
    \exp\{-k_s  (t-t_0) \mathfrak{I}_i\}N_i(t)\leq Q_i(t) \leq N_i(t),\hspace{0.2cm} \forall t
\end{equation}
\begin{equation}\label{eq27n}
    Q_i(t_0+T_i) \geq \exp\{-k_s T_i \mathfrak{I}_i\}N_i(t_0+T_i)
\end{equation}
If Assumption 1 holds, then using (\ref{eql4}) and (\ref{eq27n}) we can write
\begin{equation}
    Q_i(t_0+T_i)\geq  \exp\{-k_s T_i \mathfrak{I}_i\}\gamma_i I_{n(n+m)}
\end{equation}
When $\mathfrak{I}_i(t)=1$ then $Q_i(t)=Q(t)$, hence the following will also hold
\begin{equation}
    Q(t_0+T_i)\geq  \exp\{-k_s  T_i\}\gamma_i I_{n(n+m)}>0
\end{equation}
Implying that $Q(t_0+T_i)$ is PD.
On the other hand, $Q\left(t_{0}+T_i\right)>0$ implies $Q_i(t_0+T_i)>0$ which entails that $N_i\left(t_{0}+\right.$ $T_i) \geq Q_i\left(t_{0}+T\right)>0$, implying $N(t, x)$ being IIE for subsystem $i$ with some positive $\gamma_i$.
Hence, $N(t)$ is IIE iff $Q\left(t_{0}+T_i\right)$ is PD.\\
\textit{\textbf{Lemma 2}}: If $Q(t_{0}+T_i)$ is PD for $i\in \mathbf{S}$, $Q(t)$ will remain PD in any finite interval starting from $t=t_{0}+T_i$, i.e. $Q(t)>0 \hspace{0.1cm}\forall \hspace{0.1cm} t \in [t_{0}+T_i,t_f] \cap \bar{t}_i,$ for any $t_{0}+T_i<t_f<\infty$,
where $t_0+T_i,\hspace{0.1cm}t_f\in \bar{t}_i$ and $\bar{t}_i=\{t \hspace{0.1cm} | \hspace{0.1cm} \sigma(t)=i \}$.\\
\textbf{Proof} : From (\ref{eql1}), it can be claimed that 
\begin{equation} \label{l39}
N_{ i}\left(t_{2}\right) \geq N_{ i}\left(t_{1}\right), \text { for } t_{2}>t_{1} \geq t_{0}
\end{equation}
Let $\mathcal{H}(v, t)  \triangleq  v^{T} Q_i(t) v, \hspace{0.2cm} \forall v \in \mathbb{R}^{n(n+m)\times n(n+m)}$, 
using (\ref{l34}) and (\ref{l39}), the  following derivation is possible.
$$
\begin{aligned}
\mathcal{H}(v, t) & \geq \exp \left\{-k_s t \mathfrak{I}_i \right\} v^{T} N_{ i}(t_0+T_i) v,\hspace{0.2cm} \forall t \geq t_{0}+T_i \\
& \geq \exp \left\{-k_s t \mathfrak{I}_i \right\} \gamma_i\|v\|^{2} \\
& \geq \underbrace{\exp \left\{-k_s t_{f} \mathfrak{I}_i \right\} \gamma_i\|v\|^{2}}_{\mathcal{G}(v)>0,\hspace{0.1cm} \forall\|v\| \neq 0},\hspace{0.2cm} \forall t \leq t_{f}
\end{aligned}
$$
\\
Lemmas 1-2 indicate that the IIE condition on $N(t)$ for any $i\in \mathbf{S}$ can be verified online by checking the determinant of $Q(t)$ online; a positive value implying that the IIE condition on $N(t)$ for active subsystem is satisfied.\\
\textbf{\textit{Remark 6}}: Proof of Lemmas 1-2 are inspired from \cite{roy2017uges}. The indicator function in the IIE definition is time-varying and, therefore complicates the proofs of Lemmas 1-2. Certain modifications in the proof are made such that Lemmas 1-2 hold in the switched system context. 
\subsection{Parameter Estimation Design for Online Identification of Switched Affine Systems}
The parameter estimation law for subsystem $i\in \mathbf{S}$ is proposed as
\begin{equation}\label{eq_para_est}
    \dot{\hat{\theta}}_{i}=\begin{cases}\Gamma_{\theta i}\underbrace{(T_{li}+T_{lli}+{{s_{i}}}{T_{sw i})}}_{T_{\tilde{\theta}_{i}}},\hspace{1.3cm} \text{when} \hspace{0.2cm} \sigma(t)=i\\
    
    \Gamma_{\theta i}\underbrace{(\closure{T}_{li}+\closure{T}_{lli}+{s_{i}}T_{sw i})}_{\closure{T}_{\tilde{\theta}_{i}}},\hspace{1.3 cm} \text{when} \hspace{0.2cm} \sigma(t)\neq i
    \end{cases}
\end{equation}
\begin{equation*}
    \hspace{4 cm}\hat{\theta}_i(t_0)=\hat{\theta}_{i0},\hspace{0.1cm} \forall i \in \mathbf{S},\hspace{0.1cm} t\geq t_0
\end{equation*}
where $\Gamma_{\theta i} \in \mathbb{R}^{n(n+m)\times n(n+m)}$ is a positive-definite learning gain matrix. The terms in (\ref{eq_para_est}) are given by
\begin{subequations}
\begin{align}
    &T_{li} \triangleq k_{li} N^T(t)[g(t)-N(t)\hat{\theta}_{i}(t)] \label{eq21}&\\
    &T_{lli}\triangleq k_{lli}[G(t)-Q(t)\hat{\theta}_{i}(t)] \label{eq22}&\\
    &T_{swi}\triangleq k_{swi}[\mathbf{S_{\closure{G}}}_i-\mathbf{S_{\closure{Q}}}_i\hat{\theta}_{i}(t)]\label{eq_s}& \\
    &\closure{T}_{li}\triangleq k_{li} \mathbf{S_{N}}_i^T[\mathbf{S_{g}}_i-\mathbf{S_{N}}_i\hat{\theta}_{i}(t)]\label{eq21p}&\\
    &\closure{T}_{lli}\triangleq k_{lli}[\mathbf{S_{G}}_i-\mathbf{S_{Q}}_i\hat{\theta}_{i}(t)]\label{eq22p}&
\end{align}
\end{subequations}
\hspace{6cm}$\forall t \geq t_0,\forall i\in \mathbf{S}$\\
The piecewise-constant signal ${s_{i}}(t)\in \mathbb{R}$ is defined as
\begin{equation}\label{eq26}
    {s_{i}}(t)=\begin{cases}
    0 &\hspace{0.5cm} \text{for}\hspace{0.5cm} t\in [t_{0},\hspace{0.1cm}t_{0}+T_{i}) \\
    1& \hspace{0.5cm} \text{else}
    \end{cases}
\end{equation}
\hspace{6cm} $\forall i \in \mathbf{S}$\\
where $k_{li}, k_{lli}, k_{swi}>0$ are scalar gains and $\mathbf{S_{\closure{G}}}_i\in \mathbb{R}^{n(n+m)},\hspace{0.1cm}\mathbf{S_{\closure{Q}}}_i \in \mathbb{R}^{n(n+m) \times n(n+m)}$ and $\mathbf{S_{s}}_i \in \mathbb{R}$ denote the $i^{th}$ element of memory stack $\mathbf{S_{\closure{G}}},\hspace{0.1cm}\mathbf{S_{\closure{Q}}}$ and $\mathbf{S_s}$ corresponding to that $i^{th}$ subsystem $(i\in \mathbf{S}$). $s_i(t)=0$ implies IIE does not hold for $i^{th}$ subsystem and $s_i(t)=1$ implies IIE holds for $i^{th}$ subsystem. The resulting memory stacks are defined as $\mathbf{S_{\closure{G}}}=[\mathbf{S_{\closure{G}_1}},\mathbf{S_{\closure{G}_2}},...\mathbf{S_{\closure{G}_M}}],\hspace{0.1cm} \mathbf{S_{\closure{Q}}}=[\mathbf{S_{\closure{Q}_1}},\mathbf{S_{\closure{Q}_2}},...\mathbf{S_{\closure{Q}_M}}]$ and $\mathbf{S_s}=[\mathbf{S_{s_1}},\mathbf{S_{s_2}},...\mathbf{S_{s_M}}]$, $\mathbf{S_{s}}_i=1$ indicates IIE condition for subsystem $i \in \mathbf{S}$ is achieved. The memory stacks are populated using the following logic
\begin{subequations}
\begin{align}
&\mathbf{S_{\closure{G}}}_i  \triangleq G(t_{0}+T_{i})&\\
&\mathbf{S_{\closure{Q}}}_i  \triangleq Q(t_{0}+T_{i}) \label{eq_Q_bar}&\\
&\mathbf{S_{s}}_i   \triangleq s_{i}(t)\hspace{1.5 cm} i\in \mathbf{S} \label{eq_S_s}&
\end{align}
\end{subequations}
\textit{\textbf{Remark 7}}: The proposed adaptive identification framework facilitates parameter learning during both active and inactive periods of a subsystem. When a subsystem is active, the parameter estimation law uses composite adaptation terms $T_{li} \hspace{0.1cm} \& \hspace{0.1cm} T_{lli}$ in (\ref{eq21}), (\ref{eq22}) based on the current filter states, combined with the IIE-based term $T_{swi}$ in (\ref{eq_s}) based on the recorded filter states at time $T_i$ in the memory stacks $\mathbf{S_{\closure{Q}}},\mathbf{S_{\closure{G}}}$. $T_{swi}$ switches ON after the IIE condition for the corresponding subsystem is satisfied \cite{roy2017uges}. The parameter learning during the inactive period of a subsystem uses composite adaptation terms $\closure{T}_{li} \& \hspace{0.1cm} \closure{T}_{lli}$ in (\ref{eq21p}), (\ref{eq22p}) based on the recorded filter states in the memory stacks $\mathbf{S_N},\mathbf{S_g},\mathbf{S_Q},\mathbf{S_G}$ that carry information about the corresponding unknown parameter vector $\theta_i$.\\
\textit{\textbf{Remark 8}}: Filters in first and second layers store the values of state and input, hence only switch-out instant values are sufficient to use in the parameter estimation law in (\ref{eq_para_est})(when $\sigma(t)\neq i$) to continue parameter learning, unlike \cite{kersting2014concurrent}, where multiple data point recording of state and input is required when the subsystem is active.\\ 
\textit{\textbf{Remark 9}}: The proposed approach utilizes $M$ different parameter estimators, one for each subsystem, that are all implemented in parallel, which has certain advantages over using a single estimator for all subsystems. An important advantage of the proposed framework is that the parameter learning does not stop for inactive subsystems. Also, the implementation of the proposed switched estimator is possible since knowledge of $T_{i}$ can be obtained online by checking the determinant of $Q(t)$ and $\sigma(t)$ is assumed to be known at every time instant. The moment the determinant of $Q(t)$ becomes positive, indicating that the IIE condition for that subsystem is met, the switching term $T_{swi}$ is turned ON in the estimation law.
\section{STABILITY ANALYSIS}
The parameter estimation error is defined as
\begin{equation}
    \tilde{\theta}_{i}(t)\triangleq \hat{\theta}_{i}(t)-\theta_{i},\hspace{0.2cm}i\in \mathbf{S}
\end{equation}
Using (\ref{eq_para_est}), the dynamics of the parameter estimation error can be expressed as
\begin{equation}\label{eq27}
    \dot{\tilde{\theta}}_{i}(t)=
    \begin{cases}-\Gamma_{\theta_i}(k_{li}N^T\underbrace{N\tilde{\theta}_{i}}_{\varepsilon^T_{i}(t)}+k_{lli}Q\tilde{\theta}_{i}+s_{i}k_{swi}\mathbf{S_{\closure{Q}}}_i\tilde{\theta}_{i}), \\ \hspace{4cm} \sigma(t)=i,i\in \mathbf{S}\\
    
    -\Gamma_{\theta_i}(k_{li}\mathbf{S_{N}}_i^T\underbrace{\mathbf{S_{N}}_i\tilde{\theta}_{i}}_{\mathbf{{S_{\varepsilon}}}^T_{i}(t)}+k_{lli}\mathbf{S_{Q}}_i\tilde{\theta}_{i}+s_{i}k_{swi}\mathbf{S_{\closure{Q}}}_i\tilde{\theta}_{i}),\\\hspace{4cm} \sigma(t) \neq i,i\in \mathbf{S}
    
    \end{cases}
\end{equation}
where $\varepsilon_{i}(t)\in \mathbb{R}^{n(n+m)}$ is typically known as the prediction error \cite{slotine1989composite} and $\mathbf{S_{\varepsilon}}_i(t)\in \mathbb{R}^{n(n+m)}$ is the memory prediction error.\\
\textit{\textbf{Theorem 1}}. For the system in (\ref{eq1}), the parameter estimation law (\ref{eq_para_est}) ensures that the origin of the error dynamics $\tilde{\theta}_i(t)$ is uniformly stable. In addition, if Assumption 1 holds, the parameter estimation error $\tilde{\theta}_i(t)$ is UGES (in the delayed sense) for $t\geq t_0+T_i$, i.e.
\begin{equation}\label{eqc5}
    ||\tilde{\theta}_i(t)||\leq  \gamma_1 ||\tilde{\theta}_i(t_0+T_i)||\exp\{-\gamma_{2} (t-t_0-T_i)\},\\ \hspace{0cm} \forall t \geq t_0+T_i
\end{equation}
for some positive scalars $\gamma_1$ and $ \gamma_{2}$ independent of initial conditions, provided the following gain condition is satisfied.
\begin{equation} \label{eqc2}
    k_{swi}\lambda_{min}(\mathbf{S_{\closure{Q}}}_i)\geq \bar{\eta}_i
\end{equation}
where $\lambda_{min}(\bullet)$ denotes the minimum eigenvalue of the argument matrix, and the scalar $\bar{\eta}_i>0$ is free parameter, used to alter the rate of convergence.\\
\textbf{Proof}: Consider the following Lyapunov function candidates
\begin{equation}\label{eq38}
    V_{i}=\frac{1}{2}\tilde{\theta}_{i}^T\Gamma^{-1}_{\theta i}\tilde{\theta}_{i},\hspace{0.6cm} \forall i \in \mathbf{S}
\end{equation}
where $\Gamma_{\theta i}$ is a positive definite matrix. The Lyapunov function candidates in (\ref{eq38}) satisfy the following inequality.
\begin{equation}\label{eqc3}
    \frac{1}{2}\lambda_{mi}||\tilde{\theta}_{i}||^2\leq V_{i}\leq\lambda_{Mi} \frac{1}{2}||\tilde{\theta}_{i}||^2,\hspace{0.6cm} \forall i \in \mathbf{S}
\end{equation}
where the positive constants $\lambda_{mi}$, and $\lambda_{Mi}$ are defined as
\begin{subequations}
\begin{align}
    \lambda_{mi}&\triangleq \lambda_{min}(\Gamma^{-1}_{\theta_i})\\
    \lambda_{Mi} &\triangleq \lambda_{max}(\Gamma^{-1}_{\theta_i})
    \end{align}
\end{subequations}
where $\lambda_{min}(\bullet)$ and $\lambda_{max}(\bullet)$ denotes minimum and maximum eigenvalue of the argument matrix, respectively. Taking the time derivative of (\ref{eq38}) along the system trajectories yields
\begin{equation}\label{eqc1}
    \dot{V}_{i}=\tilde{\theta}_{i}^T\Gamma^{-1}_{\theta i}\dot{\tilde{\theta}}_{i}, \hspace{1.5cm} \forall i \in \mathbf{S}
\end{equation}
Using (\ref{eq27}) in (\ref{eqc1}), yields
\begin{equation}\label{eq41}
    \dot{V}_{i}=
    \begin{cases}-k_{li}\underbrace{\tilde{\theta}^T_iN^T}_{\varepsilon_i(t)}N\tilde{\theta}_i-k_{lli}\tilde{\theta}_i^TQ\tilde{\theta}_i-s_{i}k_{swi}\tilde{\theta}^T_{i}\mathbf{S_{\closure{Q}}}_i\tilde{\theta}_{i}\\ \hspace{ 4cm} \sigma(t)=i,\forall i \in \mathbf{S}\\
    
    -k_{li}\underbrace{\tilde{\theta}^T_i\mathbf{S_{N}}_i^T}_{\mathbf{S_{\varepsilon}}_i(t)}\mathbf{S_{N}}_i\tilde{\theta}_i-k_{lli}\tilde{\theta}_i^T\mathbf{S_{Q}}_i\tilde{\theta}_i-s_{i}k_{swi}\tilde{\theta}^T_{i}\mathbf{S_{\closure{Q}}}_i\tilde{\theta}_{i}\\
    \hspace{ 4cm} \sigma(t)\neq i,\forall i \in \mathbf{S}
    \end{cases}
\end{equation}
Depending on whether the IIE condition holds for the $i^{th}$ subsystem at the time instant $t$, two cases are possible\\
\textit{\textbf{Case 1}}: When $t<t_0+T_i$ (i.e. Before satisfying IIE for subsystem $i$) :\\
Using property 1, $\dot{V}_i$ can be upper bounded as
\begin{equation}\label{eq42}
    \dot{V}_{i}\leq
    \begin{cases}-k_{li}||\varepsilon_i(t)||^2,  \hspace{1cm} \forall t < t_0+T_i,\sigma(t)=i, \forall i \in \mathbf{S}\\
    
    -k_{li}||\mathbf{S_{\varepsilon}}_i(t)||^2, \hspace{0.8cm} \forall t < t_0+T_i,  \sigma(t)\neq i, \forall i \in \mathbf{S}
    \end{cases}
\end{equation}
From (\ref{eq42}), $\dot{V}_i$ is negative semi-definite and hence, the parameter estimation error $\tilde{\theta}_i(t)$ is bounded $\forall t <t_0+T_i$.\\
\textit{\textbf{Case 2}}: When $t\geq t_0+T_i$ (i.e. After satisfying IIE for subsystem $i$) :\\
This case indicates that Assumption 1 holds, therefore, for $t\geq t_{0}+T_i, \dot{V}_i$ in (\ref{eq41}) can be upper bounded using Lemma 2, and (\ref{eq_Q_bar}) along with the gain condition (\ref{eqc2}) as
\begin{equation}
    \dot{V}_{i}  \leq \begin{cases}-k_{li}||\varepsilon_i(t)||^2-k_{lli}\text{exp}\{-k_s (t-t_0)\}\gamma_i||\tilde{\theta}_i||^2\\\hspace{1.6cm}-\bar{\eta}_i||\tilde{\theta}_i||^2, \forall t\geq t_0+T_i,  \sigma(t)=i, \forall i \in \mathbf{S} \\
    -k_{li}||\mathbf{S_{\varepsilon}}_i(t)||^2-\eta_i ||\tilde{\theta}_i||^2-\bar{\eta}_i||\tilde{\theta}_i||^2\nonumber\\
    \hspace{3.2cm}\forall t\geq t_0+T_i, \sigma(t)\neq i, \forall i \in \mathbf{S} 
    \end{cases}\\
\end{equation}
where
    $\eta_i\leq k_{lli}\lambda_{min}(\mathbf{S_{{Q}}}_i)$. Since $Q(t)$ is a positive definite function of time after the IIE condition is satisfied, $\lambda_{min}(\mathbf{S_{{Q}}}_i)\neq 0$, from Lemma 1-2.\\
\begin{equation}
    \dot{V}_i   \leq \begin{cases} -\bar{\eta}_i||\tilde{\theta}_i||^2, \hspace{2cm} \sigma(t)=i, \forall i \in \mathbf{S}\\
    -\underbrace{(\eta_i+\bar{\eta}_i)}_{\xi_i}||\tilde{\theta}_i||^2, \hspace{0.95cm} \sigma(t)\neq i, \forall i \in \mathbf{S}
    \end{cases}
\end{equation}
Using (\ref{eqc3}), the above inequality modifies to
\begin{subequations}
\begin{align}
    \dot{V}_i & \leq
    \begin{cases}-\frac{2\bar{\eta}_i}{\lambda_{Mi}}V_i,\hspace{2cm} \sigma(t)=i\\
    -\frac{2\xi_i}{\lambda_{Mi}}V_i, \hspace{2cm}\sigma(t)\neq i
    \end{cases}\\
\dot{V}_i & \leq \begin{cases} 
 -\alpha_i V_i, \hspace{0.5cm} \forall t\geq t_0+T_i,\sigma(t)=i\\
  -\beta_i V_i, \hspace{0.5cm} \forall t\geq t_0+T_i,\sigma(t)\neq i
 \label{eq47}
\end{cases}
\end{align}
\end{subequations}
where $\alpha_i=\frac{2\bar{\eta}_i}{\lambda_{Mi}}$ and $\beta_i=\frac{2\xi_i}{\lambda_{Mi}}$ , using the Comparison Lemma (Lemma 3.4 of \cite{khalil2002nonlinear}), the differential inequality in (\ref{eq47}) leads to the subsequent exponentially convergent bound
\begin{equation}\label{eqc9}
    V_i(t)\leq \begin{cases} V_i(t_0+T_i)\exp\{-\alpha_i(t-t_0-T_i)\}\\ \hspace{3.5cm} \forall t \geq t_0+T_i, \sigma(t)=i\\
    V_i(t_0+T_i)\exp\{-\beta_i(t-t_0-T_i)\}\\ \hspace{3.5cm} \forall t \geq t_0+T_i, \sigma(t)\neq i
    
    \end{cases}
\end{equation}
Using (\ref{eqc3}), the inequality in (\ref{eqc9}) can be converted to
\begin{equation}\label{eqc10}
    ||
    \tilde{\theta}_i(t)||\leq \begin{cases} \gamma_1||\tilde{\theta}_i(t_0+T_i)||\exp\{-\gamma_{\alpha}(t-t_0-T_i)\}\\ \hspace{3cm} \forall t \geq t_0+T_i,\sigma(t)=i\\
    \gamma_1||\tilde{\theta}_i(t_0+T_i)||\exp\{-\gamma_{\beta}(t-t_0-T_i)\}\\ \hspace{3cm} \forall t \geq t_0+T_i,\sigma(t)\neq i
    \end{cases}
\end{equation}
where $\gamma_1=\sqrt{\frac{\lambda_{mi}}{\lambda_{Mi}}},\hspace{0.1cm}\gamma_{\alpha}=\frac{\alpha_i}{2},\hspace{0.1cm} \gamma_{\beta}=\frac{\beta_i}{2}$ and $\gamma_2=min (\gamma_{\alpha}, \gamma_{\beta})$ ($\gamma_2$ is used in theorem statement (\ref{eqc5})). Since, the Lyapunov function in (\ref{eq38}) is radially unbounded and the constants $\gamma_1, \gamma_{\alpha}$ and $\gamma_{\beta}$ are independent of initial conditions, the algebraic inequality in (\ref{eqc10}) proves UGES (in a delayed sense) of the parameter estimation error $\tilde{\theta}_i(t), \forall t\geq t_0+T_i$.
\section{CONCLUSION}
This paper proposes an online adaptive identification algorithm for MIMO switched affine systems, without knowledge of the system matrices and the state derivative information under the assumption of a known switching signal. The inclusion of memory allows parameter learning during inactive periods of a subsystem. A new notion of IIE condition is introduced that is shown to be sufficient for parameter convergence of switched affine system. An analytical proof that covers both stability and exponential parameter convergence of the proposed algorithm is given. An interesting future work is control design of the linear switched system with the improved parameter convergence.

\bibliographystyle{ieeetr}
{\footnotesize
\bibliography{root}}
\end{document}